\documentclass[twocolumn,showpacs,preprintnumbers,amsmath,amssymb,prl]{revtex4}

\pdfoutput=1

\usepackage{graphicx}
\usepackage{dcolumn}
\usepackage{bm}

\begin{document}

\preprint{APS/123-QED}

\title{Independent magnetization behavior of a ferromagnetic \\metal/semiconductor hybrid system}

\author{S. Mark, C. Gould, K. Pappert, J. Wenisch, K. Brunner, G. Schmidt, and L.W. Molenkamp}
 \affiliation{Physikalisches Institut (EP3), Universit\"at W\"urzburg,\\ Am Hubland, D-97074 W\"urzburg, Germany}
\date{\today}

\begin{abstract}

We report the discovery of an effect where two ferromagnetic
materials, one semiconductor ((Ga,Mn)As) and one metal (permalloy),
can be directly deposited on each other and still switch their
magnetization independently. We use this independent magnetization
behavior to create various resistance states dependent on the
magnetization direction of the individual layers. At zero magnetic
field a two layer device can reach up to four non-volatile
resistance states.

\end{abstract}


\maketitle

Devices whose functioning is based on the relative magnetization
state of two controllable magnetic elements, such as GMR (giant
magneto resistance) \cite{BAIBICH1988} \cite{BINASCH1989} based read
heads \cite{Theis2003} and TMR (tunnel magneto resistance)
\cite{JULLIERE1975} based MRAM \cite{Akerman2005} are crucial to the
modern information technology industry. So far, all such devices
have been comprised of at least three layers: the two magnetic
layers and a spacer layer to break the direct coupling between them
and allow them to reorientate their relative magnetization. In this
letter we show that, unlike the case of two ferromagnetic (FM)
metals, the bringing together of a FM metal with a FM semiconductor
(SC) can allow the layers to remain magnetically independent and
thus permit the fabrication of devices without the need of a non
magnetic interlayer. We demonstrate a first such device, which
because of the strong anisotropies in the FM semiconductor layer has
not only two, but up to four stable resistance states in the absence
of a magnetic field.

\begin{figure}
\includegraphics{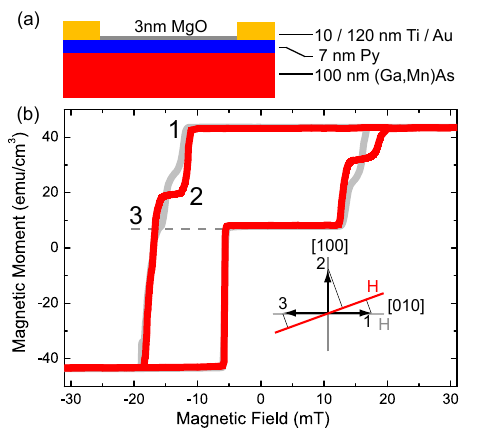}
\caption{\label{fig:squid}(a) layer stack of the hybrid system with
Ti/Au contacts. (b) Magnetization hysteresis loop of such a layer
system measured along $0^{\circ}$ (light gray) and $20^{\circ}$
(red) relative to the (Ga,Mn)As [010] easy axis after cooling the
sample to 4.2 K in a field of 300 mT. The measurements exhibit clear
double-step switching of the (Ga,Mn)As layer and a shifted
magnetization contribution of the Py/MgO system due to exchange
bias. Inset: projection of the magnetization reversal process for
the (Ga,Mn)As layer.}
\end{figure}

To prepare these structures, a 100 nm (Ga,Mn)As layer is grown by
low-temperature molecular beam epitaxy on a GaAs buffer and
substrate. Subsequently, without breaking the vacuum, the sample is
transferred to a UHV magnetron sputtering chamber, and a permalloy
(Ni$_{80}$Fe$_{20}$) film with a thickness of 7 nm (and in some
cases a 3 nm thick magnesium oxide (MgO) capping film) is deposited
on top of the (Ga,Mn)As layer (fig. \ref{fig:squid}a). Using optical
lithography and chemically assisted ion beam etching (CAIBE), this
layer stack is patterned into a 40 $\mu$m wide Hall bar oriented
along the (Ga,Mn)As [010] crystal direction. Ti/Au contacts are
established through metal evaporation and lift-off.

For an initial study of the layer system, we include an MgO film on
top of the permalloy layer to prevent the permalloy from naturally
oxidizing in air. Without a cap layer the natural oxide consisting
of NiO on an Fe oxide layer \cite{Fitzsimmons2006} produces an
exchange bias coupling \cite{MEIKLEJOHN1956} with a magnetic field
dependent anisotropy  \cite{Stiles1999} \cite{Nogues1999} below its
N\'{e}el temperature \cite{HAGEDORN1967} which unduly complicates
the layer characterization. We find that sputtering MgO on Py
creates a well-defined uniform antiferromagnetic layer which couples
antiferromagnetically to the Py film. Cooling the Py/MgO system from
above its N\'{e}el temperature to 4.2 K in an applied magnetic field
results in a fixed exchange bias coupling which is not affected by
further magnetic field sweeps. This exchange bias coupling induces a
stable unidirectional anisotropy in the Py film. Both, the
unidirectional anisotropy of Py and the principally biaxial in-plane
easy axes of (Ga,Mn)As \cite{Sawicki2004}, can be observed in direct
magnetization measurements. Figure \ref{fig:squid}b shows two
magnetization hysteresis loops of a layer system composed of
(Ga,Mn)As and Py/MgO measured by SQUID (superconducting quantum
interference device). In this experiment a magnetic field of +300 mT
has been applied during the cooling procedure from 150 to 4.2 K
along the field sweep direction. As a result, the magnetization
contribution of the Py/MgO system is shifted along the field axis
generally in the opposite ('negative') direction of the cooling
field. In addition to the Py/MgO contribution one can see the very
characteristic double-step reversal process of the (Ga,Mn)As layer.
This layer is not exchange biased, and its behavior is symmetric
around the origin. These two statements can be validated from a
detailed analysis of the hysteresis curves as follows.

In Fig. 1b, the red curve was obtained by sweeping the field along
$20^{\circ}$ with respect to the [010] (Ga,Mn)As crystal direction.
The measurement begins at +300 mT, with the magnetization of both
layers pointing along the field direction. As the field is reduced
the Py magnetization $\textbf{M}_{Py}$ continues to point in the
$20^{\circ}$ direction, whereas the (Ga,Mn)As magnetization
$\textbf{M}_{SC}$ gradually relaxes to the [010] easy axis (see
inset in fig. \ref{fig:squid}b). In our configuration, the SQUID
measures only the projection of the total moment onto the field
axis, therefore the (Ga,Mn)As magnetization rotation towards the
[010] axis changes the value to $M_{SC}cos(20^{\circ})$. This
rotation occurs at fields greater than 30 mT and is not visible in
the figure. As the field is lowered through zero, at -12 mT a
$90^{\circ}$ domain wall nucleates and propagates through the
(Ga,Mn)As layer, causing a $90^{\circ}$ switch in the direction of
its magnetization to the [100] crystal direction. At -16 mT a second
$90^{\circ}$ (Ga,Mn)As domain wall nucleates and propagates,
completing the reversal. Right after this second (Ga,Mn)As event,
the Py changes its magnetization (at -17 mT) from $20^{\circ}$ to
the $200^{\circ}$ direction. For the back sweep of the magnetic
field, because it is exchange biased, the Py layer reverses its
magnetization before zero field at -6 mT. $\textbf{M}_{SC}$ is
hysteretically symmetric and reverses its direction at positive
fields through the same double-step switching process as before.
This behavior is characteristic of the two layers responding
independently to the applied magnetic field.

After warming the sample to 150 K, and recooling with a magnetic
field along $0^{\circ}$, the second hysteresis loop with a field
sweep along the (Ga,Mn)As [010] easy axis is measured (fig.
\ref{fig:squid}b, gray curve). Due to its unidirectional anisotropy,
which is set by the exchange bias and is once again oriented along
the measurement axis, the permalloy shows a behavior identical to
the $20^{\circ}$ direction measurement. Because the sweep direction
is now along a (Ga,Mn)As easy axis, the projection onto the field
axis after the first $90^{\circ}$ switching event is almost zero,
and the two switching events occur at almost the same field with an
intermediate state having a value of half the (Ga,Mn)As total
moment. This again confirms the independent character of the two
layers.

\begin{figure}
\includegraphics{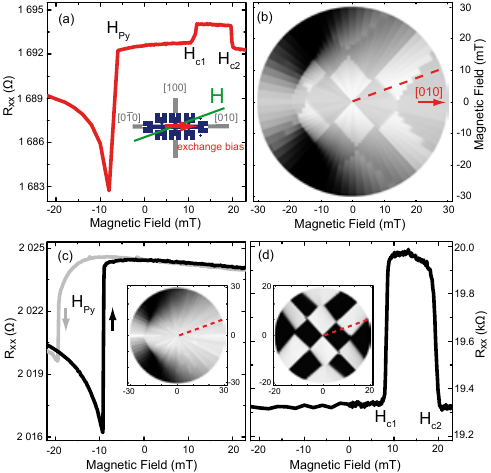}
\caption{\label{fig:polars} AMR measurements along $20^{\circ}$ for
layers composed of (Ga,Mn)As/Py/MgO (a), and control samples of
Py/MgO (c) and (Ga,Mn)As (d). b) Grayscale resistance polar plot of
the (Ga,Mn)As/Py/MgO Hall bar, with the $20^{\circ}$ direction is
marked by a dashed line. RPP for the two control layers are given in
the insets of (c) and (d).}
\end{figure}

For a more detailed analysis of this independent switching behavior,
transport measurements are performed at 4.2 K in a magnetocryostat
fitted with three orthogonal Helmholtz coils which can produce a
magnetic field of up to 300 mT in any direction. Results discussed
here are for longitudinal resistance ($R_{xx}$) measurements. The
change in $R_{xx}$ in response to an external magnetic field can be
ascribed to the anisotropic magnetoresistance (AMR)
\cite{MCGUIRE1975} \cite{Jan57}, which shows a typical
$cos^2\vartheta$-dependence where $\vartheta$ denotes the angle
between magnetization and current. In (Ga,Mn)As, the resistance
perpendicular to the magnetization is larger than the resistance
parallel to the magnetization \cite{Baxter2002} while the opposite
is true for permalloy.

Figure \ref{fig:polars}a presents a magnetoresistance curve along
$20^{\circ}$ for the hybrid system of fig. \ref{fig:squid} (100 nm
(Ga,Mn)As, 7 nm Py and 3 nm MgO), obtained after cooling the sample
in a 300 mT field oriented along $0^{\circ}$, and where one can
identify the properties of the individual layers. The reorientation
of $\textbf{M}_{Py}$ starts at negative fields and switches to the
preferred axis at $H_{Py}\sim-9 $ mT. The two (Ga,Mn)As switching
fields $H_{c1}$ and $H_{c2}$ are at $\sim11$ mT and $\sim20$ mT,
respectively. To outline the full anisotropy, the positive field
half of such magnetoresistance curves for multiple directions (here
every $5^{\circ}$) are merged into a gray scale resistance polar
plot (RPP) \cite{Pappert2007a} with the magnetic field H along the
radius as shown in fig. \ref{fig:polars}b. The gray scale encodes
the resistance values, where black denotes minimum and white maximum
resistance. The single curve from fig. \ref{fig:polars}a is along
the dashed line in fig. \ref{fig:polars}b. In order to more clearly
understand the anisotropy pattern of this hybrid system, we now
present data on characteristic individual (Ga,Mn)As and Py/MgO
layers.

Figure \ref{fig:polars}c shows AMR of a 7 nm thick Py layer capped
by a 3 nm thick MgO layer. Its primary anisotropy is unidirectional
due to exchange bias of $\sim 15$ mT. After a field cooling
procedure the unidirectional anisotropy points along the Hall bar
which is defined as the $0^{\circ}$ direction in this case. At high
negative magnetic fields along $20^{\circ}$ the Py magnetization is
antiparallel to the $20^{\circ}$ direction. As the field is brought
back towards zero (black curve) $\textbf{M}_{Py}$ rotates towards
the unidirectional easy direction. Since this is more than
$90^{\circ}$ from the original direction, this rotation initially
increases $cos^2\vartheta$, and this leads to a decrease in
resistance. At $H_{Py}\sim-9$ mT $\textbf{M}_{Py}$ switches abruptly
to the $0^{\circ}$ direction before having reached the point where
$\textbf{M}_{Py}$ and \textbf{I} are perpendicular to each other. At
zero magnetic field $\textbf{M}_{Py}||\textbf{I}$, and we observe a
high resistance state. As the magnetic field is increased further
$\textbf{M}_{Py}$ gradually rotates to the $20^{\circ}$ direction of
the external magnetic field. A back trace from high positive field
to negative field for the $20^{\circ}$ direction is also shown in
fig. \ref{fig:polars}c (light gray). For a pure unidirectional
anisotropy one expects two identical MR-curves. The deviations of
the two directions comes from an additional biaxial anisotropy in
the Py/MgO system \cite{Michel1998} with a strength of approximately
7 mT. The inset shows a set of magnetic field scans along many
angles compiled into a RPP.

Figure \ref{fig:polars}d presents a (Ga,Mn)As magnetoresistance
curve along $20^{\circ}$. At -20 mT the magnetization has already
relaxed to the $[0\overline10]$ (Ga,Mn)As easy axis. A first abrupt
resistance change at the field $H_{c1}$ happens due to a
reorientation of $\textbf{M}_{SC}$ towards the [100] (Ga,Mn)As easy
axis. A second reorientation of $\textbf{M}_{SC}$ towards [010] at
$H_{c2}$ completes the magnetization reversal. Again, a RPP is
presented in the inset.

By comparing to the RPP of the individual layers, it is clear to see
that in the RPP of the hybrid system the characteristic square
pattern of the (Ga,Mn)As anisotropies is superimposed on the pattern
from the Py layer. This behavior again demonstrates the independent
switching of the two magnetic layers in the sample.

\begin{figure}
\includegraphics{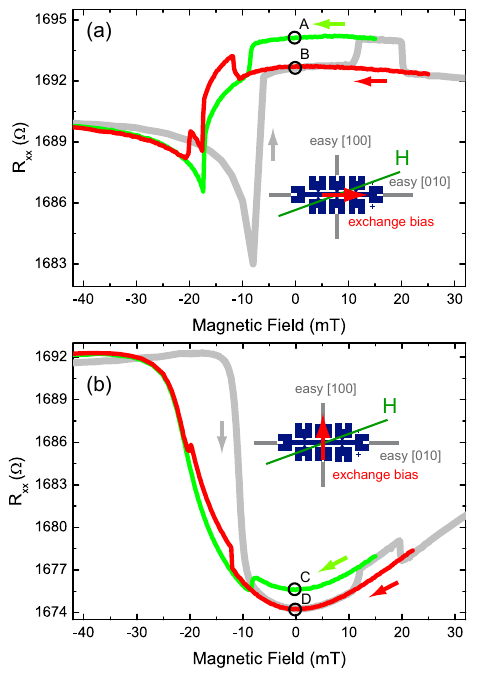}
\caption{\label{fig:minors}Minor loops measured along $20^{\circ}$
with respect to the (Ga,Mn)As [010] crystal direction on a hybrid
system. (a) with Py/MgO unidirectional anisotropy along [010] and
(b) with Py/MgO unidirectional anisotropy along [100]. light gray
reference curve, green back sweep from 15 mT and red back sweep from
25 mT.}
\end{figure}

As in GMR and TMR devices, these two independent magnetic layers
allow us to set up two non-volatile resistance states. We
demonstrate this with minor loops, sweeping the magnetic field from
negative saturation to a positive field value and back to negative
saturation. Figure \ref{fig:minors}a shows two minor loops along
$20^{\circ}$ and a reference full trace (light gray) as described in
fig. \ref{fig:polars}a. The unidirectional anisotropy in the Py
layer is now set by field cooling along the (Ga,Mn)As [010] crystal
direction. Each minor loop is of course identical to the reference
curve until their reversal point. Therefore only the back sweeps of
the minor loops are shown. Coming from negative saturation the
magnetization states of both layers are antiparallel at zero field (fig. \ref{fig:minors}a).
Sweeping the magnetic field further, the (Ga,Mn)As magnetization
reorients at $\sim11$ mT to the [100] crystal direction through a
domain wall nucleation and propagation. On stopping the field at
$\sim15$ mT and sweeping it back to zero, $\textbf{M}_{SC}$ points
perpendicular to $\textbf{M}_{Py}$. This corresponds to a high
resistance state associated with an angle of $\vartheta_{Py} =
0^{\circ}$ between \textbf{I} and $\textbf{M}_{Py}$ and an angle of
$\vartheta_{SC} = 90^{\circ}$ between \textbf{I} and
$\textbf{M}_{SC}$. Repeating the whole procedure and sweeping the
magnetic field to $\sim25$ mT instead of $\sim15$ mT, (Ga,Mn)As
completes the magnetization reversal through a second domain wall
nucleation and propagation. Back at zero field the magnetization
states of the Py and (Ga,Mn)As layers are aligned parallel to
each other. This corresponds to a high resistance state for
permalloy and a low resistance state for (Ga,Mn)As.

Figure \ref{fig:minors}b shows a similar configuration except the
unidirectional anisotropy is reset by warming the sample to 150 K
and cooling it with an appropriate applied magnetic field, to point
along [100] instead of [010]. The magnetic field is again swept in
the $20^{\circ}$ direction. At zero field the permalloy
magnetization is always parallel to [100], which is equal to a
permalloy low resistance state ($\textbf{I}\bot\textbf{M}_{Py}$).
The behavior of the (Ga,Mn)As layer is identical to the minor loop
described in fig. \ref{fig:minors}a. There are again two different
resistance states which can be ascribed to the AMR effect of the
individual layers. The minimum in fig. \ref{fig:minors}b belongs to
$\textbf{M}_{Py}$ along [100] and either $\textbf{M}_{SC}$ along
[010] or $[0\bar{1}0]$ (labeled D). At C both magnetization states
are pointing parallel to [100]. This corresponds to a high (Ga,Mn)As
resistance and a low permalloy resistance value.

To confirm that the magnetic independence of the two layers does not
originate from an electrical decoupling at the interface, we
determined the interface resistance in samples where current is
passed through the interface. The resulting contact resistance is
less than $10^{-5} \Omega cm{^2}$, which is comparable to high
quality ohmic contacts on (Ga,Mn)As. Nevertheless, we suggest that
the most plausible explanation for the lack of magnetic coupling
between the layers stems from the fact that the magnetism in the Py
layer is mediated by free electrons whereas in the (Ga,Mn)As it is
hole mediated. We note that this hypothesis is not inconsistent with
the observation that charge transport takes place freely through the
interface. For a charge current to flow between an n-type and a
p-type layer, a mechanism is only required to provide charge
conversion at the interface. In contrast, the transport of magnetic
order through the interface has much stricter requirements,
necessitating the two type of carriers to coherently exchange spin
information.

\begin{figure}
\includegraphics{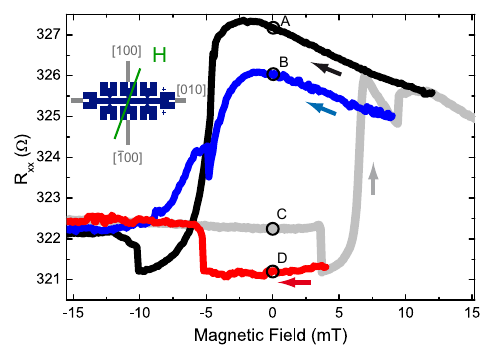}
\caption{\label{fig:4state}Magneto resistance measurements along
$70^{\circ}$ of a hybrid system composed of 70 nm (Ga,Mn)As and 7 nm
permalloy without cap layer. Minor loops starting at negative
saturation and sweeping to 4 mT, 9 mT and 12 mT.}
\end{figure}

Having established the characteristics of the two independent
layers, we now show how it can lead to a multi-value memory element.
We proceed with a new sample; a layer stack as in fig.
\ref{fig:squid}a, but the MgO left out in favor of a natural oxide
on the Py layer. This allows the use of a magnetic field to modify
\cite{Stiles1999} the exchange bias coupling direction, allowing
measurements for various Py magnetization directions at zero field,
while remaining at constant temperature. Figure \ref{fig:4state}
shows three minor loops and a full magnetoresistance curve (light
gray) of a hybrid system consisting of a 70 nm (Ga,Mn)As and a 7 nm
Py layer. Again only the back sweeps for the minor loops are shown.
The sample is cooled without an applied magnetic field. The current
path is patterned along a (Ga,Mn)as easy axis and the magnetic field
sweep direction is $70^{\circ}$.

The field sweep starts at -300 mT, and at 0 mT
(labeled C in fig. \ref{fig:4state}) the total resistance has a
lower intermediate state, where both magnetization states point
parallel towards the (Ga,Mn)As $[\overline100]$ which is associated
with a high resistance state for (Ga,Mn)As and a low resistance
state for Py. The first abrupt resistance change at $\sim$3.5 mT
corresponds to a $90^{\circ}$ reorientation of $\textbf{M}_{SC}$
towards the other (Ga,Mn)As easy axis. Sweeping the field from 4 mT
back to zero one reaches the lowest resistance state of the hybrid system (labeled
D) where both individual layers are in their low resistance state.
Lowering the magnetic field even more the (Ga,Mn)As layer switches
back to the $[\overline100]$ direction.
By sweeping the magnetic field past +9 mT, we make use of the
field rotation effect to realign the exchange bias field between the
Py and the natural oxide layer and therefore also realign the induced
anisotropy in the Py layer. Due to this reorientation the Py has a
different magnetization state for zero field. This leads to two
additional stable states (A,B) in the figure corresponding to
$\textbf{M}_{SC}$ pointing along [010] (B) and [100] (A). The
magnetization of the Py layer for these cases is almost parallel to the
current direction. These four states are comparable to those we
observed earlier with a thermal cycle (fig. \ref{fig:minors}). This
time however, all four states are achieved without warming the
sample, simply by proper manipulation of the levels using a magnetic
field.

In conclusion we have shown that independent magnetic behavior can
be obtained for magnetic layers of (Ga,Mn)As and permalloy in direct
contact without the need for a nonmagnetic interlayer. The
independence of the layers was confirmed both by transport
observations, and direct magnetization measurements using SQUID. We also
made use of this functionality to demonstrate a two layer hybrid multi-valued
memory element with four non-volatile configurations at zero
magnetic field.

\vspace{0.1cm}
\begin{acknowledgments}
The authors thank M. Sawicki for useful discussions and V. Hock and
T. Borzenko for help in sample fabrication. We acknowledge financial
support the EU (NANOSPIN FP6-IST-015728).
\end{acknowledgments}


\bibliographystyle{nature}
\bibliography{hybrid}

\end{document}